\documentstyle{article}
\begin{document}
\newcommand{\beq}{\begin{equation}}
\newcommand{\eeq}{\end{equation}}
\newcommand{\beqn}{\begin{eqnarray}}
\newcommand{\eeqn}{\end{eqnarray}}
\newcommand{\dpf}{\displaystyle\frac}
\newcommand{\no}{\nonumber}
\newcommand{\ep}{\epsilon}
\begin{center}
{\Large Can non-extremal Reissner-Nordstr$\ddot{o}$m black hole become
extremal by
assimulating infalling charged particle and shell ? }
\end{center}
\vspace{1ex}
\centerline{\large Bin
Wang$^{1,2}$\footnote[0]{e-mail:binwang@fudan.ac.cn},
\ Ru-Keng Su$^{2,1}$,\footnote[0]{e-mail:rksu@fudan.ac.cn}\ 
P.K.N.Yu$^3$ and E.C.M.Young$^3$}
\begin{center}
{$^1$ Department of Physics, Fudan University, Shanghai 200433, P.R.China\\
$^2$ China Center of Advanced Science and Technology (World Laboratory),
P.O.Box 8730, Beijing 100080, P.R.China \\
$^3$ Department of Physics and Materials Science, City University of Hong
Kong,
Hong Kong}
\end{center}
\vspace{6ex}
\begin{abstract}
By using the gedanken experiments suggested by Bekenstein and Rosenzweig,
we have
shown that non-extremal Reissner-Nordstr$\ddot{o}$m black hole cannot turn
into
extremal one by assimulating infalling charged particle and charged
spherical
shell
\end{abstract}
\vspace{6ex}
\hspace*{0mm} PACS number(s): 04.20.Dw, 04.70.Bw, 04.60.Kz
\vfill
\newpage
It was traditionally believed that the extremal black hole is the limiting
case of
its nonextremal counterpart, when the inner Cauchy horizon $r_-$ and outer
event horizon $r_+$ degenerate, the nonextremal black hole becomes extremal
[1,2].
At the extremal limit, it has been pointed out by many authors [3-8] that a
phase
transition and the corresponding scaling law exists.

But this traditional viewpoint has been challenged by many workers [9-14]
recently.
Based on an argument of the topological difference between the extremal and
the
nonextremal Reissner-Nordstr$\ddot{o}$m (RN) black hole, Hawking et al. [9]
shew that
the entropy of extremal RN black hole is zero and the formula of
Bekenstein-Hawking
entropy $S=A/4$, where $A$ is the surface area of the horizon, is not
valid. Ensuring
the stability against Hawking radiation, the temperature of the extremal
black hole
(EBH) is zero and be different from the case of nonextremal black hole
(NEBH).
They claimed that the entropy changes discontinuously in the extremal limit
implies that
one should regard NEBH and EBH as qualitatively different objects and a
NEBH cannot turn into EBH.
Furthermore, after calculating the timelike distance between $r_+$ and
$r_-$,
't Hooft [15] argued that there is no extremal limit. He said that the
macroscope EBHs are
physically illdefined limits, it would be a mistake to treat an extremal
black hole horizon
as just one horizon. On the other hand, especially in the string model [13]
or employing the brick wall model to calculate the entropy of scalar field
[11],
many other authors [11-14] have shown that the extremal limit still exist
and the Bekenstein-Hawking
entropy formula is still valid.
However, recently the details of what makes an extremal black hole
different from a nonextremal
black hole has also been observed as of particular importance in string
theory [16].
All these investigations indicate that the entropy and
other physical properties are still far from fully understood in the
extremal limit.

To clarify this puzzle, it is of important to study whether one can use
some physical
processes such as radiation, absorption etc. to realize the extremal limit
and make
the NEBH becomes EBH. The horizons of RN black hole locate at
\beq      
r_{\pm}=M \pm \sqrt{M^2-Q^2}
\eeq
where $M$ and $Q$ are the mass and charge of the hole respectively. For
NEBH, $M>Q$;
for EBH, $M=Q$. Obviously, if a physical process can increase the charge
$Q$ or
decrease the mass $M$ of the hole, one can realize the extremal limit and
transform the
NEBH into EBH. It was shown in [17] that the radiational process such as
Hawking
radiation, Penrose process and super-radiance cannot transform the NEBH
into an extremal one.
In addition to radiation, the black hole has another important character,
namely, absorption. Can the
tremendous gravitation of the RN black hole be used to swallow charges and
make
the NEBH becomes extremal? This is the objective that we hope to discuss in
this
paper.

In the last few years, an appealing problem whether one can push the black
hole over the brink
to violate the extremal condition and remove the event horizon was
addressed
by many authors. The first negative answer to this problem was put forward
by
Hiscock [18]. A quite similar approach of studying the problem but with
some
quantum point of view was suggested by
Bekenstein and Rosenzweig (BR)[19], in which two gedanken experiments to
investigate
the stability of the event horizon were proposed. In their reference, an
attempt was made to
violate the condition $Q^2\leq M^2$ by adding to an extremal RN black hole
entity which has
charge $q$ in such a way that $q^2+Q^2>(M+mE)^2$, where $Q\in U(1)$ and
$q\in U(1)'$,
$U(1)$ and $U(1)'$ are different gauge fields. $E$ is specific energy and
$m$
the mass of the infalling entity. The entity can be a spherical shell or a
point charge.
They addressed different kinds of processes carrying both classical and
quantum machanical
entities and studied whether this charge can enter the RN black hole and
break
the extremal condition. They found that all processes in which such
entities can be used to remove the event horizon
are forbidden. The same result was also obtained by Jensen [20] with an
attempt by
adding matter with a negative gravitational energy to reduce the mass term
of RN black hole.
In a previous paper [21], we extended the above studies
to three dimentional Banados, Teitelboim and Zanelli black holes. In this
paper, we will
employ the BR gedanken experiments to study another problem, namely,
whether the extremal condition can be realized.
We will prove that not only processes which break the extremal condition
are forbidden, but
also processes which make
\beq             
q^2+Q^2=(M+mE)^2
\eeq
and transform the NEBH into EBH are forbidden as well. A nonextremal RN
black hole
cannot  assimulate the infalling charged particle and/or charged shell to
become
an EBH from these gedanken experiments.

1.Infall of charged shell

Following [19], the exterior metric for a spherical distribution with two
different $U(1)$
charges $Q$ and $q$ is
\beq            
{\rm d}s^2=-(1-\dpf{2M}{r} +\dpf{Q^2+q^2}{r^2}){\rm d}t^2+\dpf{{\rm
d}r^2}{1-\dpf{2M}{r}+\dpf{Q^2+q^2}{r^2}}+
           r^2({\rm d}\theta^2+\sin^2\theta{\rm d}\phi^2)
\eeq
The equation of motion of a particle which locates at the outer edge of the
shell is
\beq           
(\dpf{{\rm d}r}{{\rm
d}\tau})^2-\dpf{2}{r}[M+mE(1-\dpf{q^2}{m^2})]+\dpf{1}{r^2}[Q^2+q^2(1-\dpf{q^
2}{m^2})]=E^2-1>0
\eeq
Where $E$ is fixed to $E>1$ for all particles [19]. Defining the terms
following
$(\dpf{{\rm d}r}{{\rm d}\tau})^2$ as the effective potential, we see there
is a
potential barrier outside the horizon and the maximum value
\beq              
V_{max}=\dpf{[mE(q^2/m^2-1)-M]^2}{q^2(q^2/m^2-1)-Q^2}
\eeq
is located at
\beq     
r_e=\dpf{q^2}{mE}+\dpf{Mq^2/mE-Q^2}{mE(q^2/m^2-1)-M}
\eeq
Since we begin with a nonextremal RN black hole, satisfying
\beq         
M^2>Q^2
\eeq
and hope the resulting configuration becomes extremal and satisfy Eq(2)
after
the assimulation of the charged shell. Combining Eq(2) and Eq(7), we can
write
\beq             
q^2=m^2E^2+2MmE+m^2\alpha,\,    Q^2=(1-\beta^2)M^2
\eeq
where $\alpha >0$ and $0<\beta<1$. Thus we obtain
\beq             
r_e>2M
\eeq
and
\beq           
V_{max}-(E^2-1)=m^2 \dpf{(E^2\beta -E^2-\beta)M^2-(E^2m^2-m^2+\alpha
m^2+2EmM)\alpha}{Q^2m^2-q^4+m^2q^2} >0
\eeq
Eq(9) means the shell will reach the potential barrier before reaching the
horizon.
Eq(10) means the shell must reach a turning point before reaching the
maximum of
the potential. The potential will prevent the shell from falling into the
black hole
to transform a NEBH to EBH.
\newpage
2.Infall of point charge

Suppose the point charge with mass $m$ and charge $q$ is a test particle,
its equation of motion is
\beq                 
(\dpf{{\rm d}r}{{\rm d}\tau})^2 - \dpf{2M}{r}+\dpf{Q^2}{r^2}=E^2-1>0
\eeq
because of $m\ll M, q\ll Q$. We can easily prove that in this case the
potential cannot
prevent the particle from falling into the black hole.

But we notice another mechanism. If the resulting configuration becomes
extremal
after absorbing the test particle, the charge $q$ must satisfy
\beq         
q^2=M^2+2MmE+m^2E^2-Q^2>2MmE+m^2E^2
\eeq
Similar to the classical electron radius, the classical charge radius is
\beq                
r_c=q^2/m >2ME+mE^2 >2M
\eeq
by using Eq(12). If the point charge infalls into NEBH and turn it into
EBH, its
radius must be bigger than that of the original black hole radius. By using
the same
argument of ref.[19], we conclude that the attempt of transforming NEBH
into EBH by
assimulating point charge is impossible.

In summary, employing two gedanken experiments suggested by Bekenstein and
Rosenzweig, we
examine the possiblity whether the nonextremal RN black hole can be
transformed into
the extremal one by assimulating infalling charged particle and shell. All
results are negative.
Besides Hawking radiation, Penrose process and super-radiance, the
absorption of
charged particle and shell cannot transform the NEBH into EBH. Even though
the
Bekenstein-Rosenzweig gedanken experiments cannot rule out all absorptive
processes, but as
examples, it seems to support the argument of Hawking et al. [9], EBH is a
different
object from that of NEBH and can not be developed from the NEBH.\\
\vspace{1ex}

\hspace{0mm} This work was supported in part by NNSF of China.

\end{document}